# Room Temperature Micro-Photoluminescence Studies of Colloidal WS$_2$ Nanosheets

*André Philipp Frauendorf [1,‡], André Niebur [2,3,‡], Lena Harms [4], Shivangi Shree [5], Bernhard Urbaszek [5], Michael Oestreich [1,6], Jens Hübner [1,6,†], Jannika Lauth [2,3,6,\*]*

[1]Institute of Solid State Physics, Leibniz University Hannover, Appelstr. 2, D-30167 Hannover, Germany

[2]Institute of Physical Chemistry and Electrochemistry, Leibniz University Hannover, Callinstr. 3a, D-30167 Hannover, Germany

[3]Cluster of Excellence PhoenixD (Photonics, Optics, and Engineering – Innovation Across Disciplines), Hannover, Germany

[4]Institute of Chemistry, Carl von Ossietzky University of Oldenburg, Carl-von-Ossietzky Str. 9-11, D-26129 Oldenburg, Germany

[5]Université de Toulouse, INSA-CNRS-UPS, LPCNO, 135 Avenue Rangueil, 31077 Toulouse, France

[6]Laboratory of Nano and Quantum Engineering (LNQE), Leibniz University Hannover, Schneiderberg 39, D-30167 Hannover, Germany

ABSTRACT

Wet-chemical syntheses for quasi two-dimensional (2D) transition metal dichalcogenides (TMDs) have emerged as promising methods for straightforward solution-processing of these materials. However, photoluminescence properties of colloidal TMDs are virtually unexplored due to the typically non-emitting synthesis products. In this work, we demonstrate room temperature micro-photoluminescence of delicate ultrathin colloidal $WS_2$ nanosheets synthesized from $WCl_6$ and elemental sulfur in oleic acid and oleylamine at 320 °C for the first time. Both, mono- and multilayer photoluminescence are observed, revealing comparable characteristics to exfoliated TMD monolayers and underpinning the high quality of colloidal $WS_2$ nanosheets. In addition, a promising long-term air-stability of colloidal $WS_2$ nanosheets is found and the control of photodegradation of the structures under laser excitation is identified as a challenge for further advancing nanosheet monolayers. Our results render colloidal TMDs as easily synthesized and highly promising 2D semiconductors with optical properties fully competitive with conventionally fabricated ultrathin TMDs.

Atomically thin TMDs with the chemical formula $MX_2$ (M = Mo, W; X = S, Se, Te) are at the forefront of a new generation of optically active 2D materials[1,2] that are promising candidates for single-photon emission, as ultra-fast transistors,[3] and for spin- or valleytronics[4–7]. Optical characterization methods including low-temperature spectroscopy[8] have established a profound knowledge of the unique optoelectronic properties of TMDs that are dominated by the formation of long-living excitons with high binding energies in the order of hundreds of meV.[9] In conjunction with a direct band gap in the visible to near-infrared region of the optical spectrum[10], a large spin splitting induced by a strong spin-orbit coupling[11] and a valley-selective addressable supplemental degree of freedom[12,13], TMD monolayers have the potential to impact a wide spectrum of research fields.

Most of the current experimental findings were obtained using samples fabricated via exfoliation[14–16] or chemical vapor deposition (CVD)[17–19] since these methods yield monolayers of large lateral sizes and low defect concentration. However, as an additional alternative fabrication method wet-chemical synthesis methods for TMDs are becoming increasingly popular.[20–23] These colloidal TMDs represent solution-processable and scalable materials and – if the material thickness is controlled properly – bear a high potential as building blocks in TMD heterostructures for numerous applications like photodetectors,[24] solar cells,[25] or polarized light emitting diodes.[26] Current research has shown that the approach is capable of producing various TMDs with single- to few-layer thickness and controlled lateral size.[27,28] Nevertheless, despite the huge synthetic progress, the optical properties of colloidal TMD mono- and few-layers prepared by colloidal methods have been sparsely studied[22,23,29] and photoluminescence (PL) in non-passivated ultrathin nanosheets has not been shown up to now due to competing defect-induced non-radiative mechanisms.

In this work we present the room temperature micro-photoluminescence ($\mu$-PL) spectroscopic characterization of wet-chemically synthesized WS$_2$ nanosheets, showing that WS$_2$ nanosheet mono- and multilayers exhibit similar optical quality like 2D WS$_2$ prepared by exfoliation or CVD and rendering the colloidal TMDs fully compatible with existing methods for the first time. In addition, close attention is paid to both, the promising long-term air-stability as well as to the occurring photodegradation process, which turns out to be the most challenging task for future detailed research. The findings in this work aim for the optimization of the optical properties of colloidal TMDs via a systematic investigation of their joint structural and $\mu$-PL properties.

Photoluminescent colloidal WS$_2$ nanosheets were synthesized by adjusting wet-chemical methods existing in literature for obtaining semiconducting WS$_2$ mono- and few layers.[20,30] In these studies, different crystal phases that are either semiconducting or metallic were obtained and exhibited no PL. We focused on synthetically advancing the PL properties of WS$_2$ nanosheets by producing the semiconducting phase. This included a better solubility of the tungsten precursor (WCl$_6$), which was achieved by dissolving WCl$_6$ in a mixture of oleylamine (OlAm) and oleic acid (OA). OlAm and OA form an oleylammonium species in an acid-base equilibrium[31] and can assemble as a lamellar tungsten precursor which keeps its structure at higher reaction temperatures.[32,33] In contrast to previously described WS$_2$ nanosheet syntheses, elemental sulfur dissolved in OlAm is used as less toxic alternative to CS$_2$ or H$_2$S typically used as sulfur sources. Both precursor solutions are combined and subsequently added to an OlAm and hexamethyldisilazane (HMDS) solution at 320 °C to synthesize WS$_2$ nanosheets. A highly reactive S$_4$N$_4$ species is released additionally to an alkylammonium polysulfide species that steadily releases H$_2$S at higher reaction temperatures, when elemental sulfur reacts with HMDS in oleylamine. First, a S-N polymer intermediate is formed through a reduction of the dissolved sulfur by HMDS. This S-N polymer intermediate then mainly decomposes into $S_4N_4$ rings due to the presence of the tungsten cations.[34,35]

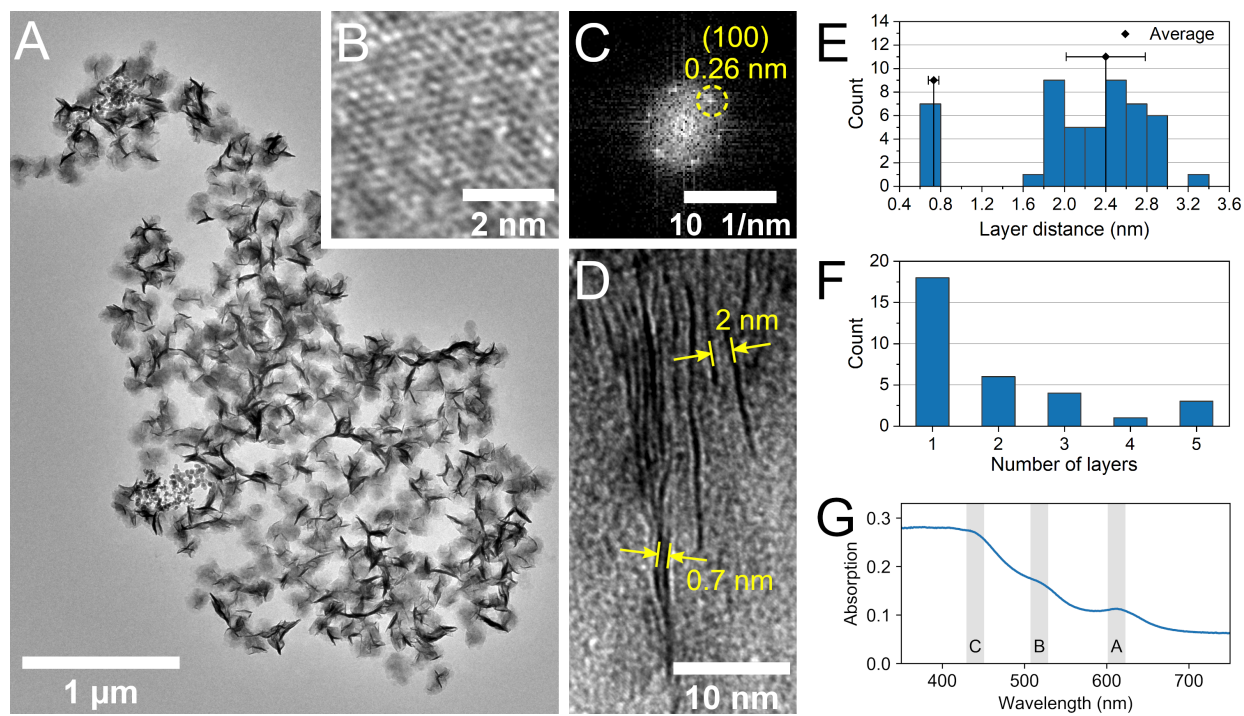

**Figure 1.** [A] Overview TEM image of colloidal $WS_2$ nanosheets with flower-like morphology. [B] HR-TEM image showing flat-lying nanosheets with a hexagonal lattice of the material confirmed by the FFT [C]. [D] TEM images revealing layer distances with and without interlayer ligands of 0.73 ± 0.06 nm and 2.4 ± 0.4 nm, respectively, as depicted in [E]. [F] shows that monolayers and few layers without ligand intercalation of up to five stacked layers are found in the sample. [G] Absorption spectrum of the as-prepared $WS_2$ nanosheets with the A-, B-, and C-excitonic transitions at 612 nm (2.03 eV), 521 nm (2.38 eV), and 441 nm (2.81 eV), respectively. In order to correct for scattering, the samples were measured inside an integrating sphere.

The enhanced reactivity of the sulfur precursor results in an accelerated crystal growth and a higher defect density. Consequently, the transition temperature of the metallic 1T- to the semiconducting 2H-crystal phase is decreased so that a higher content of the 2H crystal phase is obtained.[36] A detailed description of the synthesis conditions and chemicals is found in the Supporting Information (SI).

Figure 1A-D shows transmission electron microscopy (TEM) images of $WS_2$ mono- and multilayers. The overview (Figure 1A) demonstrates that the nanosheets form a flower-like morphology with overall lateral dimensions of a few micrometers. Figure 1B shows a high-resolution (HR)-TEM image of a flat lying domain of a single $WS_2$ nanosheet, revealing its crystallinity, while the FFT depicted in Figure 1C underpins the hexagonal periodicity and the presence of the semiconducting 2H crystal phase. The measured distance of the (100) reflex of 0.26 nm is consistent with experimental values in the literature.[27] This is also confirmed by the FFT (Figure S1B) of a larger area shown in Figure S1A and the powder XRD depicted in Figure S1C. Interlayer distances in $WS_2$ nanosheets visible by the structures stacked on their edges are shown in Figure 1D. Statistical analysis of these distances carves out two major cases shown in Figure 1E: 1) layers are 0.73 ± 0.06 nm apart, or 2) layers exhibit a less defined spacing of 2.4 ± 0.4 nm. These distances correlate well with the theoretical tungsten-to-tungsten layer distance in the bulk material of 0.62 nm[37] and an interim ligand layer of several nanometers due to a ligand templating effect, respectively.[38] As shown in Figure 1F, monolayers and few-layers of up to five stacked sheets are found by TEM. Therefore, both mono- and multilayer features can be expected in the PL measurements. Absorption spectra of $WS_2$ nanosheets exhibit a negligible background absorption at 750 nm (1.65 eV), which is attributed to small amounts of the metallic $WS_2$ phase present in the samples.[20] However, distinct absorption features are visible for the A-, B- and C-exciton at 612 nm (2.03 eV), 521 nm (2.38 eV), and 441 nm (2.81 eV), as depicted in Figure 1G, and confirm the predominant semiconducting character of the synthesized $WS_2$ nanosheets.[20] A detailed evaluation of the A-excitonic feature is shown in Figure S2 and reveals a small shoulder at 633 nm (1.96 eV) in the differentiated absorption spectrum. The presence of these two absorption energies is attributed to mono- and few layers as has been shown for CVD samples before[39] and as is validated by the TEM images.

In contrast to absorption measurements, PL characterization is substantially more influenced by defects present in the nanosheets,[8] which can reduce the PL efficiency by the stimulation of different non-radiative recombination mechanisms.[40–43] Up to now, typical colloidal TMDs exhibit a high number of chalcogen vacancies leading to a lack of PL in the samples and hindering their unambiguous optical investigation.[29] Based on the essential importance of PL measurements for future research and for underpinning the high structural quality of colloidal TMDs shown here, we apply highly sensitive $\mu$-PL spectroscopy for a systematic investigation of the optical properties of $WS_2$ nanosheets. The employed confocal microscope setup is schematically shown in Figure S3 in the SI. It contains, *inter alia*, a high working distance objective (NA = 0.45), motorized translation stages and a nitrogen-cooled CCD array, allowing spatially resolved, highly sensitive measurements with excitation and detection spot sizes in the range of 1-2 $\mu$m.

All $\mu$-PL measurements were performed at a temperature of 25 °C. Two samples were prepared by drop casting a solution with three different concentrations of the $WS_2$ nanosheets on a standard broadband dielectric mirror (Edmund Optics) substrate. Sample 1 was assembled instantaneously into a micro-cryostat while sample 2 was stored for six months under ambient conditions in a container without a desiccant for the validation of the samples' air-stability.

We find $\mu$-PL of 2D colloidal $WS_2$ mono- and multilayers, underpinning the high quality of single ultrathin $WS_2$ layers. Figure 2 shows $\mu$-PL measurements of $WS_2$ nanosheets illustrating the different aspects of the measured monolayer PL. Spectra of different positions *on* (position 2-4) and *next to* (position 1) the measured PL feature (Figure 2A) can be clearly assigned to a $WS_2$ monolayer due to the narrow linewidth and matching emission energy of the spectral features alongside with a drastic increase of the PL intensity.[10,44,45] All spectra were obtained by acquiring two PL spectra at each position which were subsequently filtered, averaged and subtracted with a background noise spectrum. The corresponding pixel of the positions are highlighted in the PL map (see Figure 2B) showing the integrated spectra. We find an elongated shape of the monolayer feature with a spatial diameter below 5 $\mu$m, which can be identified as well in the microscope image of the sample (Figure S4).

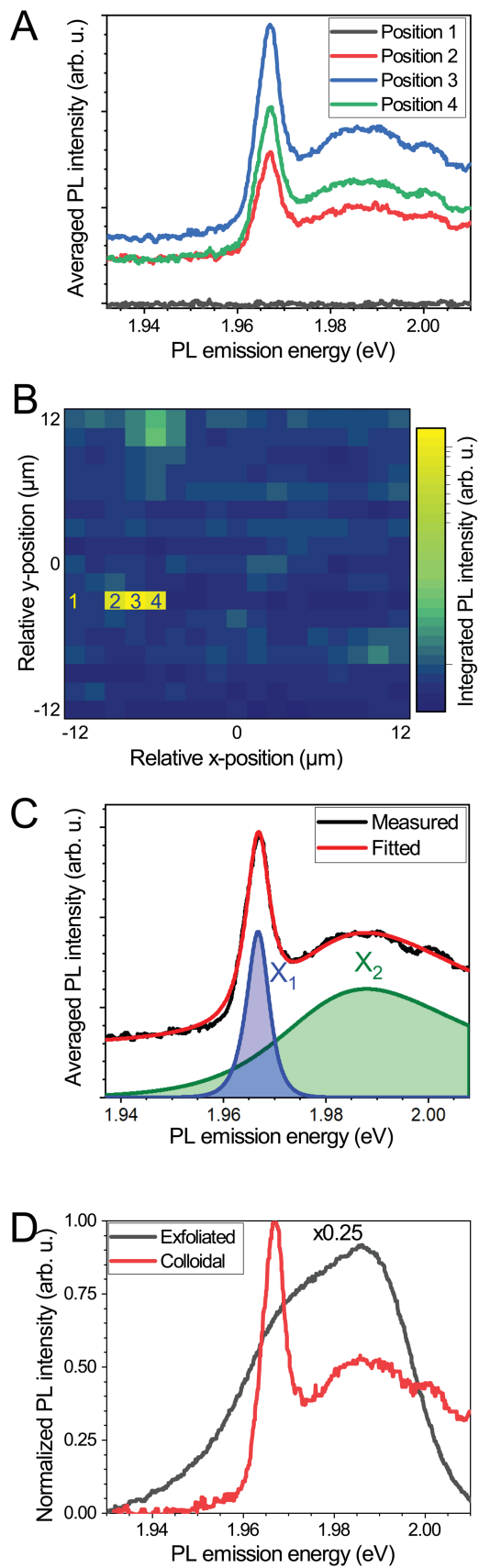

**Figure 2.** Monolayer PL of colloidal $WS_2$ nanosheets: [A] Averaged PL spectra at different positions *on* (position 2 to 4) and *next to* (position 1) the monolayer. [B] Related spatial map of the integrated PL intensity with a lateral step size of 1.5 $\mu$m. [C] Result of spectral fitting with an asymmetric hyperbolic secant function with narrow PL ($X_1$) at ~1.97 eV associated to localized excitons or trions and PL ($X_2$) associated to neutral excitons. [D] Comparison of the normalized monolayer PL signal of position 3 with the PL spectrum of an exfoliated $WS_2$ monolayer. The colloidal $WS_2$ nanosheets exhibit considerable PL intensity similar (only four times lower) to exfoliated $WS_2$ monolayers. For clarity, the offset of the colloidal PL signal was subtracted.

The PL spectrum of the colloidal $WS_2$ nanosheets is composed of three contributions (Figure 2C), each described by an asymmetric hyperbolic secant function.[46] We assign the nearly constant offset in this energy range to multilayer PL and the transition $X_2$ to the neutral exciton. The dominating narrow PL peak at ~1.97 eV ($X_1$) might be induced by localized excitons[47] or by trions formed by an electron surplus due to sulfur vacancies in colloidal $WS_2$ nanosheets.[48–52]

When comparing the normalized spectra of the colloidal $WS_2$ monolayer PL at position 3 with the PL of a state-of-the-art hexagonal boron nitride (hBN)-encapsulated exfoliated $WS_2$ monolayer (sample 3, see Figure 2D), both samples exhibit similar properties in general, with colloidal $WS_2$ nanosheets showing an only fourfold lower PL intensity as the compared exfoliated $WS_2$ sample. Surprisingly, the PL features of colloidal $WS_2$ nanosheets are much stronger pronounced than the exfoliated monolayer. Taking into account the long-standing experience in the field of the exfoliation fabrication method[8], our results show the overall promising optical properties of the wet-chemically synthesized TMDs. We find colloidal $WS_2$ nanosheet monolayers to be rather degradation sensitive (visible in the two successively taken PL spectra of position 4, see Figure S5), which could be, e.g., attributed to an optically triggered oxidation process.

In conjunction with the absorption and TEM measurements of the colloidal $WS_2$ nanosheets, a considerable number of features was found in the $\mu$-PL measurement likewise, which we assign to few-layer structures due to significantly lower PL intensities and broader widths.[10,45] These multilayer features exhibit a direct relation between microscope image and PL map (Figure 3A and Figure 3B) and are present either as separate structures (Figure 3C) or as an extended agglomerate of two or more features with overall lateral dimensions of up to tens of $\mu$m (Figure S6). In agreement with the TEM image analysis shown in Figure 1, the typical separated multilayer feature has an approximately circular shape with an overall diameter of less than $5\mu$m.

The associated broad PL spectra (see Figure 3C zoom-in) consist of multiple spectra for different central wavelengths of the spectrometer and provide the expected width of several hundred of meVs due to the influence of the indirect band-gap transition.[10,45] For future applications of colloidal $WS_2$ nanosheets it is fundamental to study their critical photodegradation and aging processes as well as to further optimize their PL characteristics (Figure S7). The understanding of the excitation power dependence of the PL intensity $I$ (Figure 3D and Figure S8) of colloidal $WS_2$ nanosheets enables a quantitative analysis of both and reveals a nearly linear dependence ($k \sim 0.91 \pm 0.01$) by fitting the data with a standard power function $I(P) \sim P^k$. This dependence is expected for the radiative recombination of excitons and trions including small deviations possibly based on mechanisms like the defect-induced formation of bound excitons or phonon side-band emission.[53,54] However, the $\mu$-PL measurements likewise reveal photodegradation under laser excitation in colloidal $WS_2$ nanosheets. For this analysis, suitable multilayer features were initially localized with PL maps and exposed to the laser in a controlled manner. A typical exponential decrease of the averaged PL intensity is shown in Figure 3E. Here, the laser excitation intensity of $\sim 1.9 \times 10^4$ W cm$^{-2}$ induces a drop of the averaged PL intensity below 50% of the initial value within ten minutes, also directly visible in the related PL spectra (Figure S9). Generally, this laser induced exponential decreases of the PL with decay times in the range of $3 \pm 0.3$ to $13.7 \pm 1.2$ minutes (Table S1) are in good agreement with photodegradation processes in exfoliated or CVD grown TMDs reported earlier.[55–58] We assume that the photodegradation of colloidal $WS_2$ nanosheets is likewise driven by a photoinduced oxidation process including chemisorption[58] or etching of the structure[57]. Additional work is required to prevent the degradation process with a promising approach being the formation of colloidal TMD heterostructures. Both lateral and stacked TMD heterostructures, like hBN supported TMD heterostructures[56], as well as core-crown geometries[26] are promising for further tailoring the optoelectronic properties of colloidal $WS_2$ nanosheets in order to prevent photodegradation.

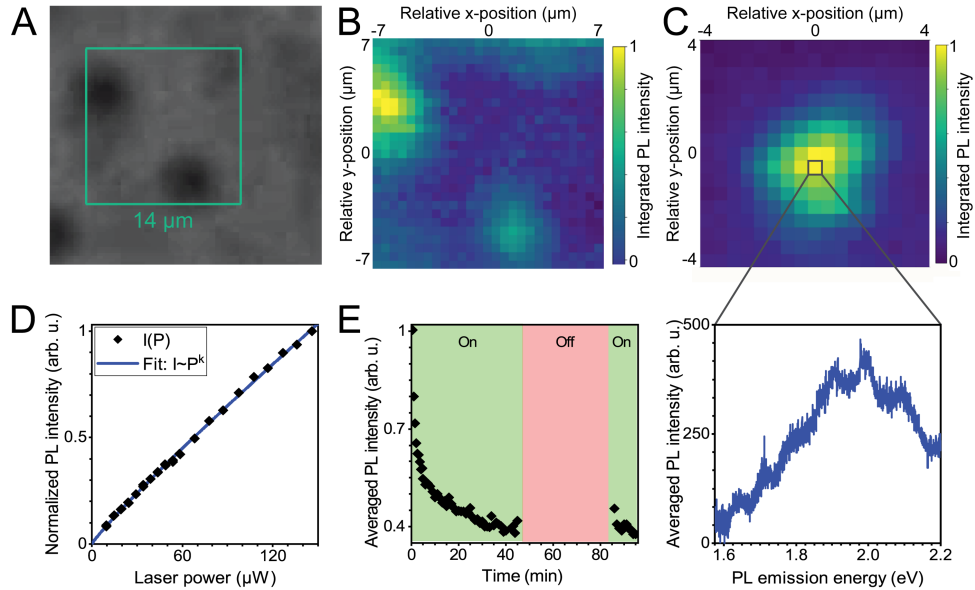

**Figure 3.** Multilayer PL of colloidal $WS_2$ nanosheets: [A] Microscope image of separated $WS_2$ multilayer features in a high concentration area of the aged sample revealing the direct relation with the associated PL map [B]. [C] Integrated PL spectral map of a separated multilayer feature with the complementary broad PL spectrum shown in the zoom-in. [D] Excitation power dependence of the $WS_2$ nanosheets multilayer PL intensity exhibiting nearly linear behavior. [E] Photodegradation induced decrease of the multilayer PL intensity. The laser excitation was set-out in the area marked red.

Generally, the degradation of the optical properties of TMDs appear primarily under elevated conditions such as laser exposure, but have been observed likewise under ambient conditions due to environmental aging processes before.[55] The air-stability of the colloidal $WS_2$ nanosheets is analyzed by comparing the directly investigated (fresh) sample 1 with the six months old (aged) sample 2. Critical PL characteristics of both samples are summarized depending on the $WS_2$ nanosheet solution concentration at drop-casting in Table S2 (including the number of multilayers per 100 $\mu m^2$ and the averaged PL intensity). We find that despite the air exposure, significant PL signals with comparable or even higher averaged PL intensities are measured in all concentration areas of the aged sample and, in conjunction with the microscope images (Figure S10), a higher number of separated multilayer features is detected. The latter might be introduced either by normal variations in the fabrication process or by stress-induced reformations, which have been observed before in $WS_2$[46]. The overall comparable PL characteristics of the aged $WS_2$ nanosheet sample and the fresh one underpin the remarkable air-stability of the colloidal $WS_2$ nanosheets investigated in this work in comparison to other studied encapsulation-free TMDs.[55,59,60]

In conclusion, we have shown mono- and multilayer room temperature $\mu$-PL of colloidal 2D $WS_2$ for the first time, underpinning the high synthetic quality of single ultrathin $WS_2$ nanosheets and revealing comparable characteristics to exfoliated $WS_2$ monolayers. Challenges for further advancing colloidal $WS_2$ nanosheet monolayers include the control of their initial photodegradation under laser excitation and are currently investigated. The observed long-term stability of the colloidal samples under ambient conditions finally advances colloidal $WS_2$ nanosheets as straightforward-to-synthesize and highly promising optical materials into a competitive level with conventionally produced 2D TMDs.

ASSOCIATED CONTENT

**Supporting Information**.

The following Supporting Information is available free of charge.

Methods, data analysis, concentration dependence of the PL characteristics, photodegradation, interference effect, HR-TEM image and XRD pattern, absorption spectrum, detailed description of the $\mu$-PL setup, microscope image of the monolayer feature, degradation of the monolayer PL, microscope image and PL map of an agglomerated multilayer feature, quantitative analysis of the concentration dependence of the multilayer PL, PL spectra of the excitation power dependence, PL spectra of the photodegradation induced PL decrease, microscope images and PL maps of the three concentration areas, intensity and position dependence of the photodegradation, characteristics of the interference effect, parameter of the exponential fit of the photodegradation induced PL decrease, multilayer PL characteristics. (PDF)

AUTHOR INFORMATION

**Corresponding Authors**

† PD Dr. Jens Hübner, jhuebner@nano.uni-hannover.de

\* Dr. Jannika Lauth, jannika.lauth@pci.uni-hannover.de

**Author Contributions**

The manuscript was written through contributions of all authors. All authors have given approval to the final version of the manuscript. ‡André P. Frauendorf and André Niebur contributed equally to this work.

The authors declare no competing financial interests.

ACKNOWLEDGMENT

J.H. and M.O. gratefully acknowledge financial support by the Deutsche Forschungsgemeinschaft (DFG, German Research Foundation) under Germany's Excellence Strategy -EXC-2123 QuantumFrontiers - 390837967 and HU 1318/4-1. J.L. greatly acknowledges funding by the Deutsche Forschungsgemeinschaft (DFG, German Research Foundation) under Germany's Excellence Strategy within the Cluster of Excellence PhoenixD (EXC 2122, Project ID 390833453). J.L. is thankful for funding by the Caroline Herschel program of the Leibniz Universität Hannover.

ABBREVIATIONS

TMD, transition metal dichalcogenide; CVD, chemical vapor deposition; PL, photoluminescence; OlAm, oleylamine; OA, oleic acid; HMDS, hexamethyldisilazane; TEM, transmission electron microscopy; hBN, hexagonal boron nitride

**Table of Contents Graphic**

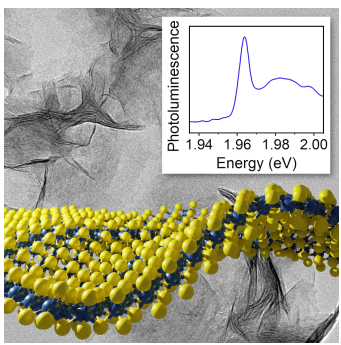

Supporting Information:

# Room Temperature Micro-Photoluminescence Studies of Colloidal WS$_2$ Nanosheets

*André Philipp Frauendorf [1,#], André Niebur [2,3,#], Lena Harms [4], Shivangi Shree [5], Bernhard Urbaszek [5], Michael Oestreich [1,6], Jens Hübner [1,6,†], Jannika Lauth [2,3,6,\*]*

[1]Institute of Solid State Physics, Leibniz University Hannover, Appelstr. 2, D-30167 Hannover, Germany

[2]Institute of Physical Chemistry and Electrochemistry, Leibniz University Hannover, Callinstr. 3a, D-30167 Hannover, Germany

[3]Cluster of Excellence PhoenixD (Photonics, Optics, and Engineering – Innovation Across Disciplines), Hannover, Germany

[4]Institute of Chemistry, Carl von Ossietzky University of Oldenburg, Carl-von-Ossietzky Str. 9-11, D-26129 Oldenburg, Germany

[5]Université de Toulouse, INSA-CNRS-UPS, LPCNO, 135 Avenue Rangueil, 31077 Toulouse, France

[6]Laboratory of Nano and Quantum Engineering (LNQE), Leibniz University Hannover, Schneiderberg 39, D-30167 Hannover, Germany

AUTHOR INFORMATION

**Corresponding Authors**

† PD Dr. Jens Hübner, jhuebner@nano.uni-hannover.de

* Dr. Jannika Lauth, jannika.lauth@pci.uni-hannover.de

**Methods:**

**Chemicals**

Oleylamine (70 %, technical grade) was acquired from Sigma-Aldrich and was degassed under vacuum for 6 h at 120 °C and stored under nitrogen. All other chemicals were used as received without further purification and stored under nitrogen. Tungsten(VI) chloride ($WCl_6$, 99%), and hexane (anhydrous) were purchased from Alfa Aesar. $WCl_6$ was stored at –25 °C. Sulfur (99.98%) and oleic acid (≥ 99%) were obtained from Sigma-Aldrich. 1,1,1,3,3,3-hexamethyldisilazane (HMDS, 98 %) was purchased from Acros Organics and kept at –25 °C.

**$WS_2$ Nanosheet Synthesis**

0.038 mmol (15 mg) $WCl_6$, 0.3 mL oleic acid (OA), and 3 mL oleylamine (OlAm) were mixed in a glovebox and ultrasonicated for 1 h in a closed vial under nitrogen. In a second vial 2.37 mmol (76 mg) of elemental sulfur and 4 mL OlAm were mixed inside a glovebox. The sulfur was dissolved stirring at 110 °C for 25 min. In agreement with previous reported $WS_2$ syntheses we found that a considerable sulfur excess for reaction (60 equivalents) supports the growth of the semiconducting crystal phase.[S1,S2] Both steps, the sonication of the tungsten precursor and the heating of the sulfur precursor, are necessary to avoid poorly dispersed starting precursors in the syringes that could clog tight cannulas. During preparation of the tungsten and sulfur precursors, 15 mL of degassed OlAm was transferred into a three-neck flask and connected to a Schlenk-line. The OlAm was degassed by heating to 85 °C for 30 min under vacuum. Subsequently, the flask was set under argon and 150 μL HMDS were added with a syringe before the mixture was heated to 320 °C. While heating, the tungsten and sulfur precursor solutions were combined in a syringe under nitrogen. Subsequently the precursor solution was added to the flask over 30 min with a syringe pump. After adding a few droplets, the clear yellow reaction mixture turned black and opaque. The reaction mixture was held at 320 °C for 30 min and cooled to room temperature afterwards.

The following precipitation and re-dissolving were performed inside a glovebox to prevent a reaction with oxygen. The synthesis product was precipitated by adding 16 mL hexane and centrifugation. The clear brownish supernatant was decanted and the $WS_2$ nanosheets were redispersed in 5 mL hexane, ultrasonicated for 5 min, and kept under nitrogen for characterization.

**Transmission Electron Microscopy (TEM)**

Samples were prepared on carbon coated copper grids and imaged with a Tecnai G2 F20 TMP from FEI. An acceleration voltage of 200 kV was applied.

**Absorption**

Absorption spectra were obtained with a Cary 5000 UV-vis-NIR spectrometer from Varian. The measurements were performed on colloidal $WS_2$ nanosheet solutions inside an integrating sphere to correct for scattering.

**Powder X-ray diffraction (XRD)**

Powder X-ray diffractograms were acquired using a Bruker D8 Advance with a $CuK_{\alpha 1}$ source at 40 kV and 30 mA. Samples were prepared on a silicon single crystal by drop-casting the concentrated product colloid.

**Micro-Photoluminescence (μ-PL)**

The μ-PL measurements were performed in reflection geometry with a homebuilt confocal microscope set-up, schematically shown in Figure S3. For the excitation a frequency-doubled Nd:YAG laser (2.33 eV) was used. The linearly polarized (LP) laser beam was reflected by a beam splitter (BS) to a long working distance objective lens (O, NA = 0.45), focusing it to the sample (S) with a typical spot size in the range of 1-2 $\mu$m. The sample itself was located under high vacuum (~3x10$^{-6}$ mbar) in a micro-cryostat. Motorized translation stages (TS) ensured an optimal positioning in x-, y- and z-direction and thereby enabled measurements with spatial resolutions on a sub-micrometer scale.

The PL signal was directed into a high-resolution Czerny-Turner spectrometer and detected with a nitrogen-cooled charge coupled device (CCD) array. A longpass filter with a transmission cut-on energy (2.25 eV) above the laser excitation energy was used to avoid negative impacts of the excitation laser on the PL spectra. For *in situ* monitoring of the sample an optional imaging tool was adapted, consisting of a removable mirror (RM), a white light source (WL) and a camera.

All μ-PL measurements were performed at a temperature of 25 °C. Based on the tungsten atom concentration in the product of $c_{\text{theor}}(W) = 7.6$ mmol/L for a neglected loss, the WS$_2$ nanosheet solution was diluted with hexane to a concentration of $c_{\text{high}}(W) = 760$ μmol/L, $c_{\text{med}}(W) = 76$ μmol/L, and $c_{\text{low}}(W) = 7.6$ μmol/L and deposited onto two standard broadband dielectric mirrors (broadband dielectric $\lambda/10$ mirror, Edmund Optics[S3]). Sample 1 was assembled in the micro-cryostat right after the synthesis. For the validation of the long-term stability, sample 2 was stored for six months under ambient conditions in a container without a desiccant before the investigation. Additionally, an exfoliated WS$_2$ monolayer sample was used for benchmarking of the results. It was fabricated on a comparable broadband dielectric mirror using an all-dry viscoelastic stamping method.[S4] An encapsulation with a thick bottom and thin top hexagonal boron nitride (hBN)-layer was applied to optimize the optical properties.[S5] Suitable hBN-flakes and the $WS_2$ monolayer were identified with a microscope based on their contrast.[S6]

**Data analysis**

All PL spectra were filtered with a median filter in order to subtract occasionally occurring discharge spikes in the spectra recorded by the high-sensitivity CCD array. The background noise was subtracted, and the spectra were subsequently averaged and processed depending on the particular measurement. For broad PL spectra, single spectra were acquired at different central wavelengths of the spectrometer and compiled to the final broad spectrum. PL maps were obtained by integrating the single spectra and constructing the map with the related positions. The characteristics of these maps were obtained *via* the manual identification of each PL feature and the following determination of the averaged PL intensity in consideration of the background noise and the laser excitation power.

**Photodegradation:**

The delicate colloidal $WS_2$ structures exhibit photodegradation under laser excitation. This photodegradation process is investigated by measuring the induced decrease of the PL intensity. Suitable positions were initially selected via PL maps and exposed with a defined laser excitation. PL spectra were acquired at chosen time steps and subsequently smoothed with a median filter. The background noise as well as the laser excitation power were monitored during the measurement to notice and subtract changes. For a quantitative analysis of the photodegradation induced decrease of the PL the averaged PL was normalized with respect to the during the maps measured original PL intensity. Significant parameters like the integration time and the laser excitation were taken into account. The decrease of the PL intensity $I$ was conclusively fitted with an exponential function:

$$I(t) = I_0 + A\, e^{-t/\tau} \qquad (1)$$

The critical parameters of the decay time $\tau$ and the constant PL intensity $I_0$ are summarized in Table S1 for the different measurements and show no pronounced difference for the directly investigated and aged sample.

However, the photodegradation induced decrease of the PL is affected by the laser excitation power (Figure S12A) and the position (Figure S12B). The intensity dependence was measured at similar positions in the medium concentration area and shows a faster and stronger decrease of the PL intensity for higher excitation power (Figure S12A). The position dependence was measured with comparable laser excitation powers at different positions of the high concentration area, revealing a varying PL decrease for these positions.

**Concentration dependence:**

For future applications of colloidal WS$_2$ nanosheets it is essential to further optimize their generally promising PL characteristic. In our study we find the drop-casting concentration of the nanosheet solution as easily controllable parameter to achieve a PL enhancement. The concentration dependence of the PL is evaluated in the aged sample based on the more uniform distribution of the multilayer features (Figure S10 and Figure S11) and reveals a lower number of PL features with increasing concentrations. Nevertheless, fitting the data with the adapted Poisson distribution:

$$N(\bar{I}) = I_0\, \frac{e^{-r}\, r^{b\bar{I}}}{(b\,\bar{I})!} \qquad (2)$$

shows that an increasing drop-casting concentration is connected with the formation of larger WS$_2$ nanosheet assemblies, inducing a shift to PL features with higher averaged PL intensities (Figure S7). In this case $N$ is the number of multilayer features per 1000 $\mu$m$^2$, $\bar{I}$ is the averaged PL intensity and $r$, $b$ and $I_0$ are the fitting parameters of the adapted Poisson distribution (Table S3). We assume this shift to be caused either by a higher probability to find an emitting feature or by a higher number of emitting features as is evident from the spacing of the layers in the TEM analysis (see Figure 1 in the main text).[S7] At last the formation of a shielding of multiple layers of the oleylamine and oleic acid ligands, directly visible in the microscope images (Figure S10), might *increase* the PL intensity comparable to an encapsulation with hBN.[S8] Our findings thereby verify that it is possible to enhance the PL characteristics of WS$_2$ nanosheets by rationally diluting the WS$_2$ nanosheet solutions and provide a starting point for future PL studies.

**Interference effect:**

A negative interference effect was observed in the PL spectra of the colloidal WS$_2$ nanosheets. To investigate this phenomenon, multiple broad PL spectra were acquired at different positions of the directly investigated sample 1 and the aged sample 2. The interference effects seem to have an influence on PL both, the spectra of the aged sample (Figure S13A) and the spectra of the directly investigated sample (Figure S13C). However, in the directly investigated sample 1, spectra with a negligible influence (Figure S13B) are present likewise. The interference effect varies between different concentration areas but is constant within a region (Figure S13C) and is likely to be related to thin film interference of same thicknesses within a region. Therefore, we assume it to be introduced by the ultrathin structure of the WS$_2$ nanosheets as has been observed before in 2D materials.[S9,S10] In addition, the formation of a capping of multiple layers of "dried" oleic acid and oleylamine ligands, directly visible in the microscope images of the high concentration (Figure S10), might influence the effect comparable to similar effects in hBN encapsulated

TMDs.[S11] Further work is needed to validate these assumptions and to find a way to suppress the interference effects based on simulations of the interference function.

**Supplemental Figures:**

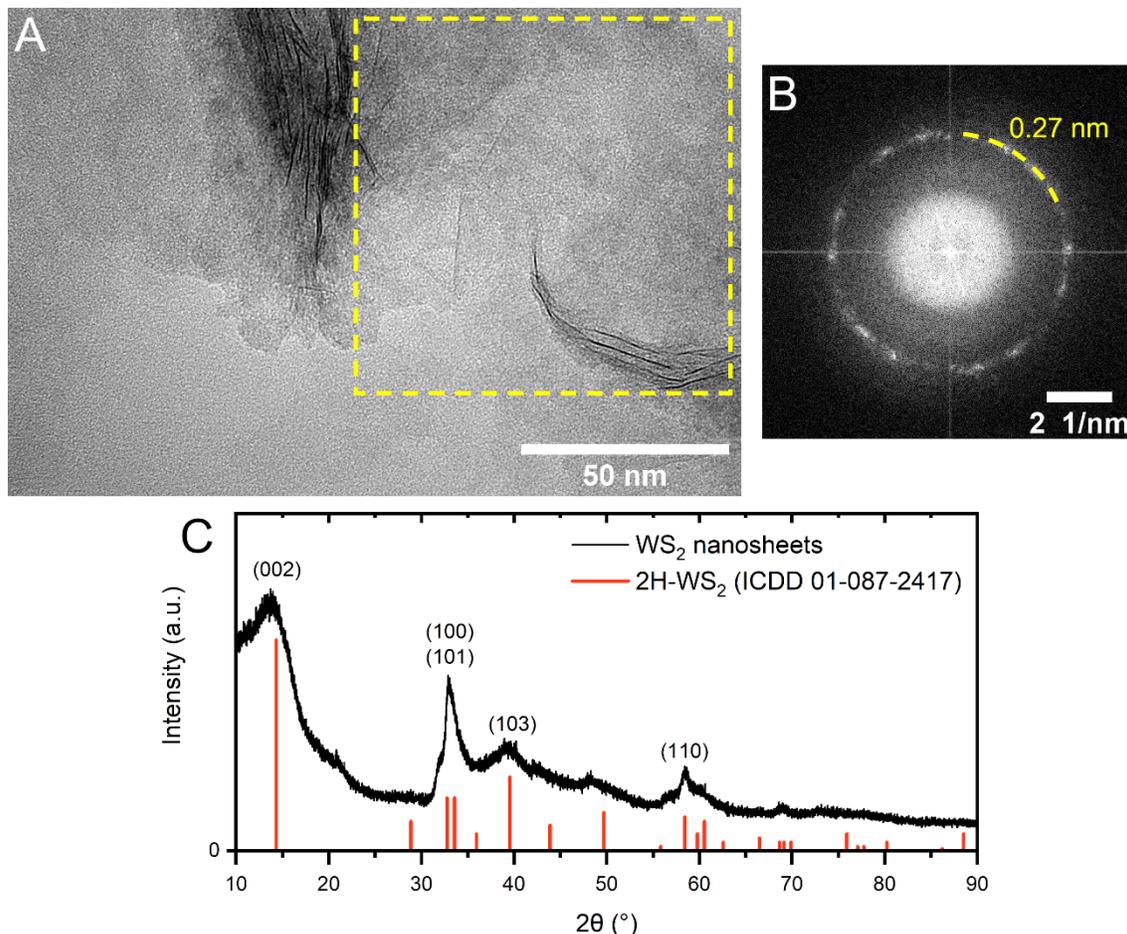

**Figure S1.** Crystallographic characterization of $WS_2$ nanosheets: [A] HR-TEM image of $WS_2$ nanosheets and [B] associated FFT of the yellow marked area in [A]. The arranged features in [B] marked by the yellow circle section correspond to the (100) lattice plane of hexagonal $WS_2$ due to the nanosheets lying flat and randomly oriented on the TEM grid. [C] Powder XRD pattern of the as-prepared $WS_2$ nanosheets. The diffractogram agrees well with the reflexes of the database reference measurement on bulk 2H-$WS_2$. Some reflexes are broadened or lack presence as is typical for 2D nanosheets and has been reported for 2H-TMD nanosheets before.[S1,S2]

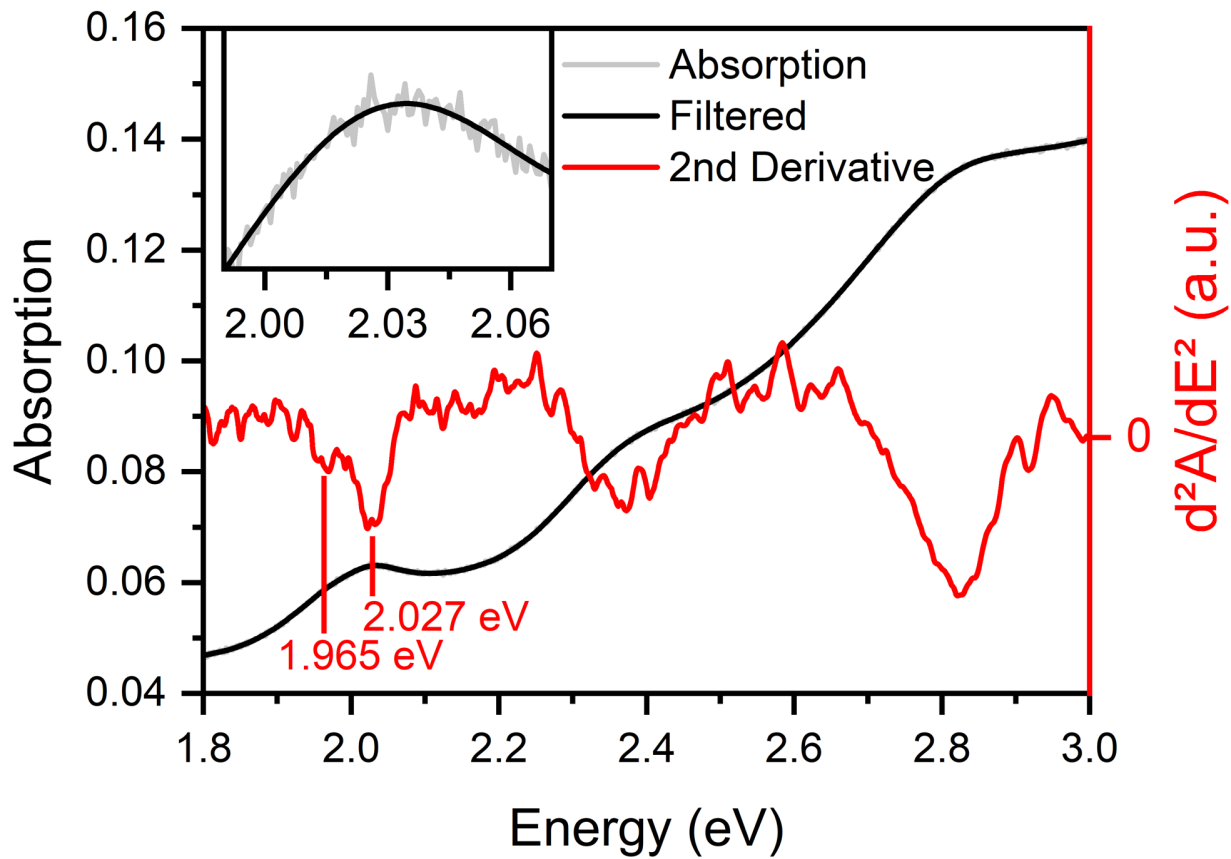

**Figure S2.** Absorption spectrum of the as-prepared colloidal $WS_2$ nanosheets measured in an integrating sphere to correct for scattering. Excitonic transitions appear in the form of shoulders or maxima. The exact values of the transitions were determined by using the 2nd derivative of the absorption to find the point of maximum curvature. Before derivation, the absorption data were filtered by local regression (*LOESS filter*).

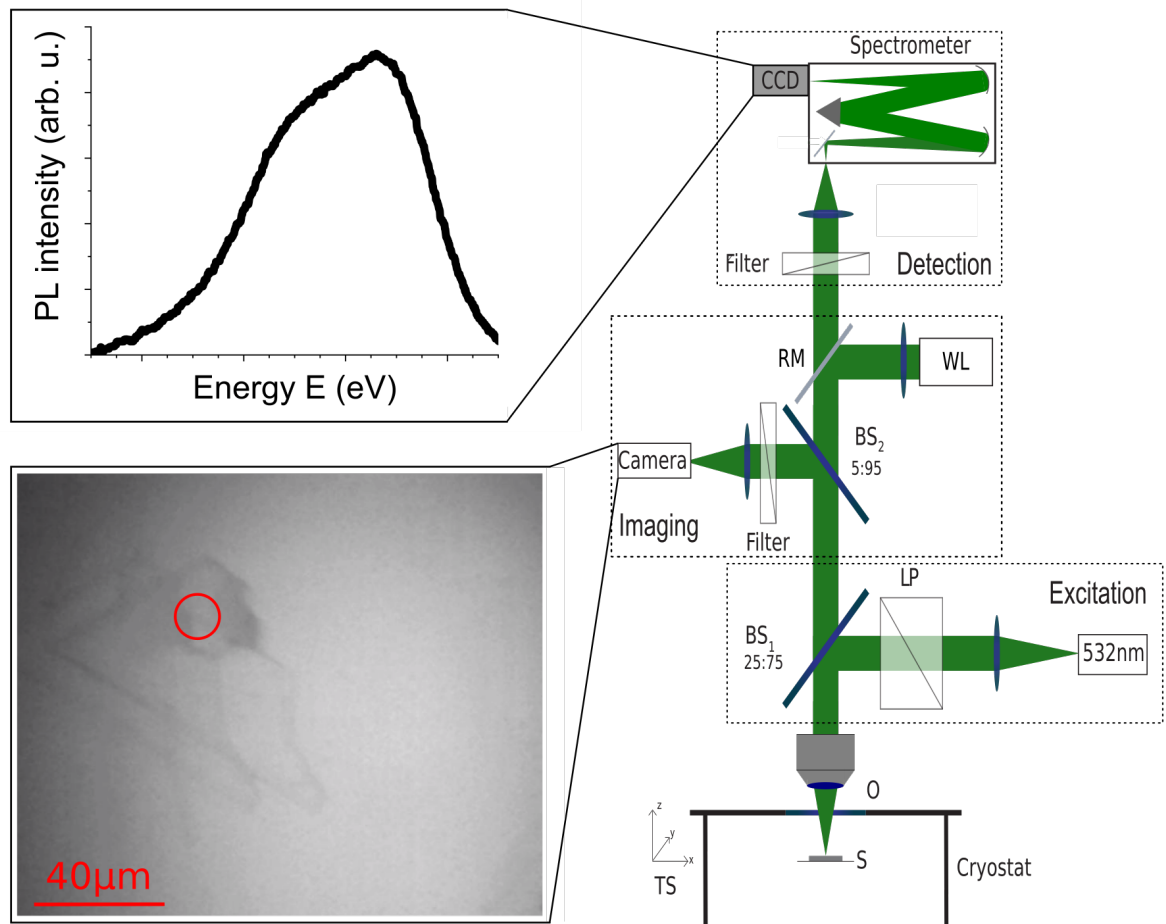

**Figure S3.** Simplified schematic of the $\mu$-PL setup with an exemplary PL spectrum and microscope image of the exfoliated and hBN encapsulated WS$_2$ monolayer. The red circle highlights the monolayer position.

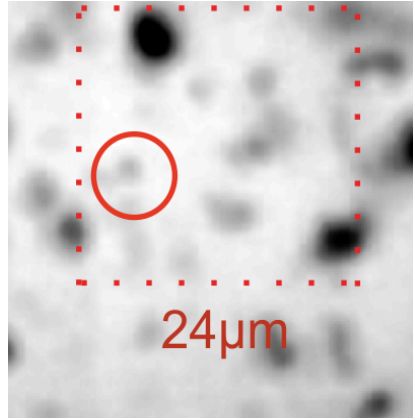

**Figure S4.** Magnified microscope image of sample 1 close to the monolayer feature. The position of the monolayer was calculated based on the laser spot position and is highlighted with the red circle. In addition, the red square indicates the area of the associated PL map (Figure 2B).

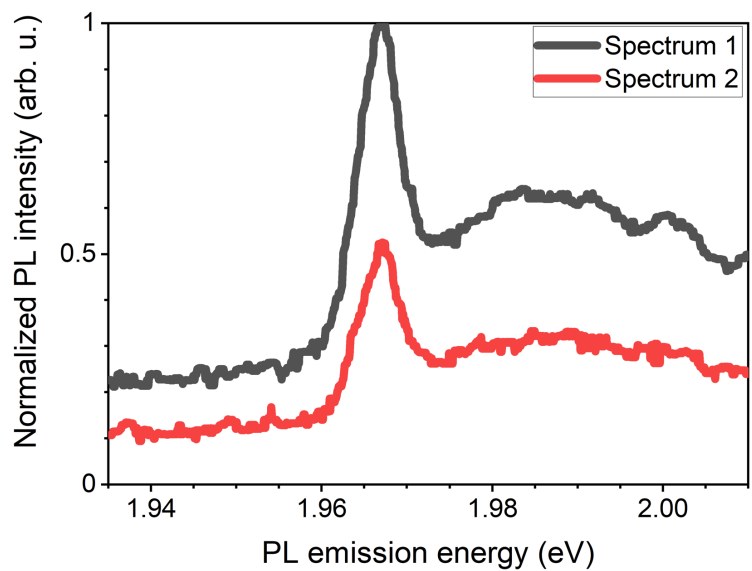

**Figure S5**. Consecutively acquired single PL spectra of colloidal $WS_2$ nanosheets showing the degradation of the monolayer PL signal at position 4 of the related PL map (Figure 2B).

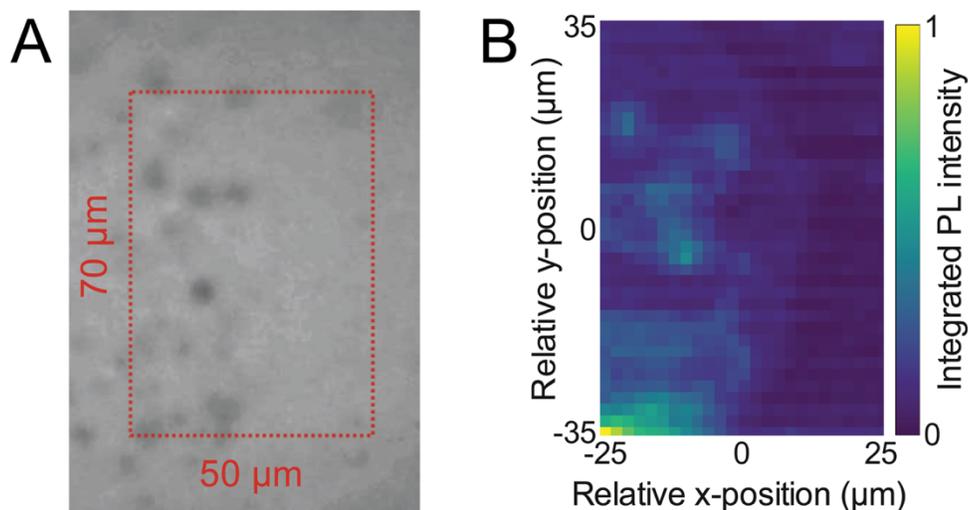

**Figure S6.** [A] Enlarged microscope image and related spatial map of the integrated PL intensity [B] of an extended agglomerate of colloidal WS$_2$ nanosheets in the medium concentration area of sample 1. These extended agglomerates occur especially in the medium and high concentration areas due to the stochastic preparation process.

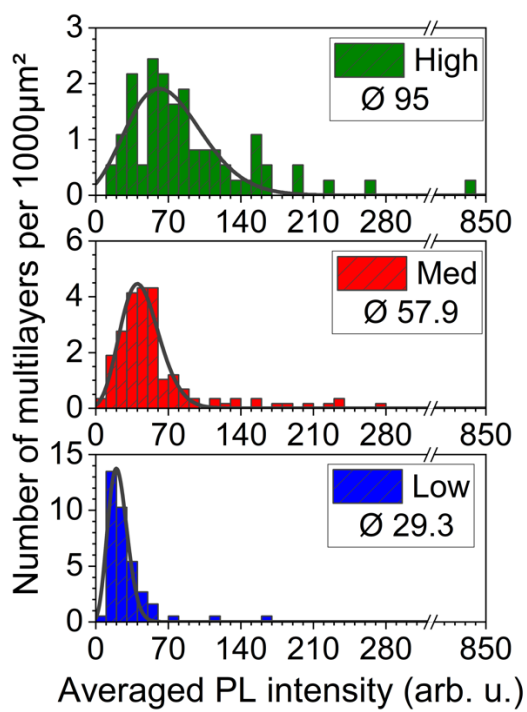

**Figure S7.** Quantitative analysis of the WS$_2$ nanosheet multilayer PL (aged sample) for high, medium and low drop-casting concentrations. The specified mean values of the PL intensity were obtained by averaging the multilayer PL intensities of each concentration.

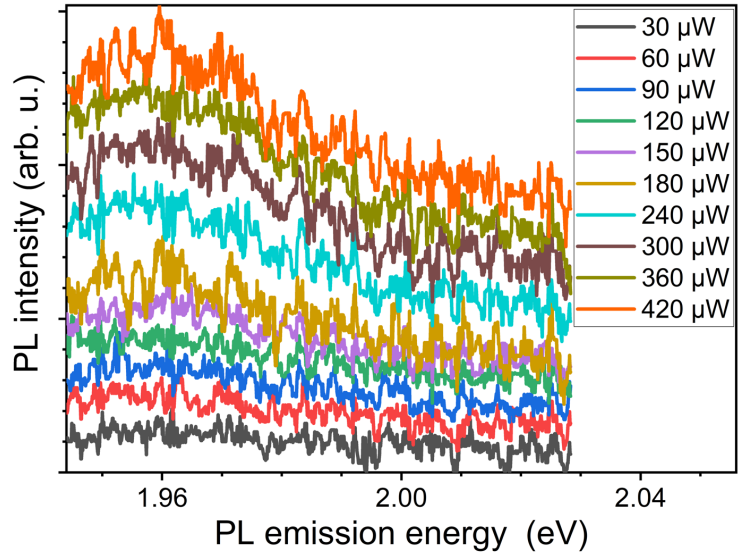

**Figure S8.** PL spectra of the excitation power dependence of the multilayer PL shown in Figure 3D and discussed in the main text.

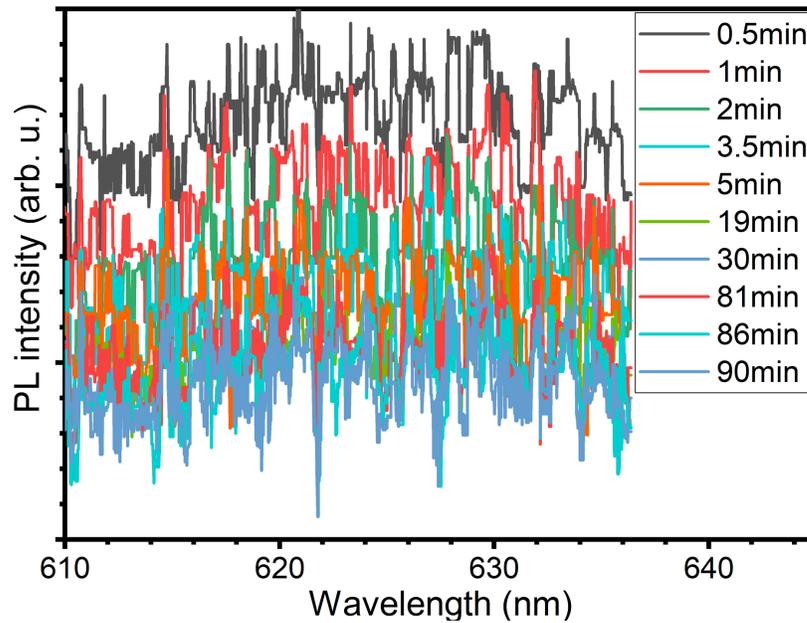

**Figure S9.** Selected single spectra of the photodegradation induced decrease of the PL shown in Figure 3E. A laser excitation power of 155 $\mu$W was used.

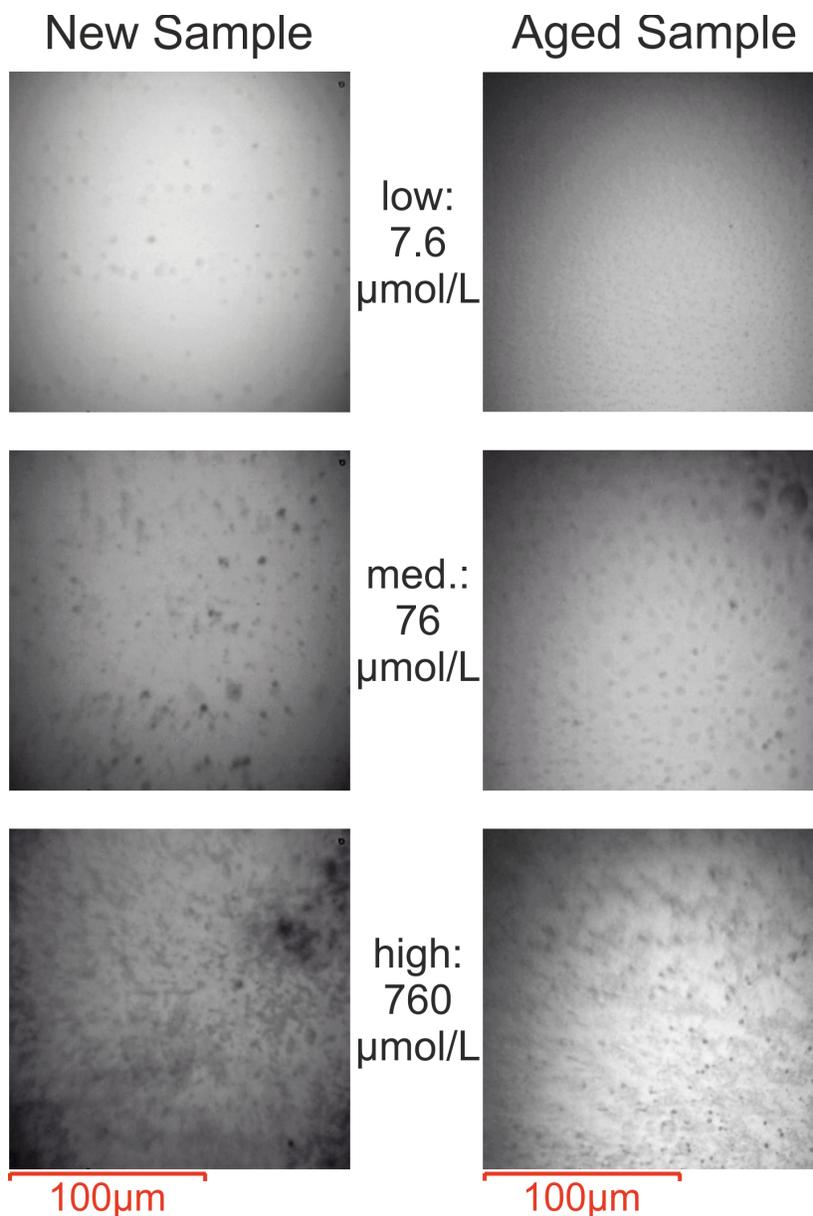

**Figure S10.** Typical microscope images for three concentrations of the directly investigated (fresh) and six months old (aged) sample. The different concentrations were obtained by diluting the precipitated product solution after fabrication and were used to investigate the influence of the concentration on the optical properties. It should be noted that the final allocation of the colloidal WS$_2$ nanosheets exhibits a stochastic nature which results in small variations that might occur within areas of the same concentration.

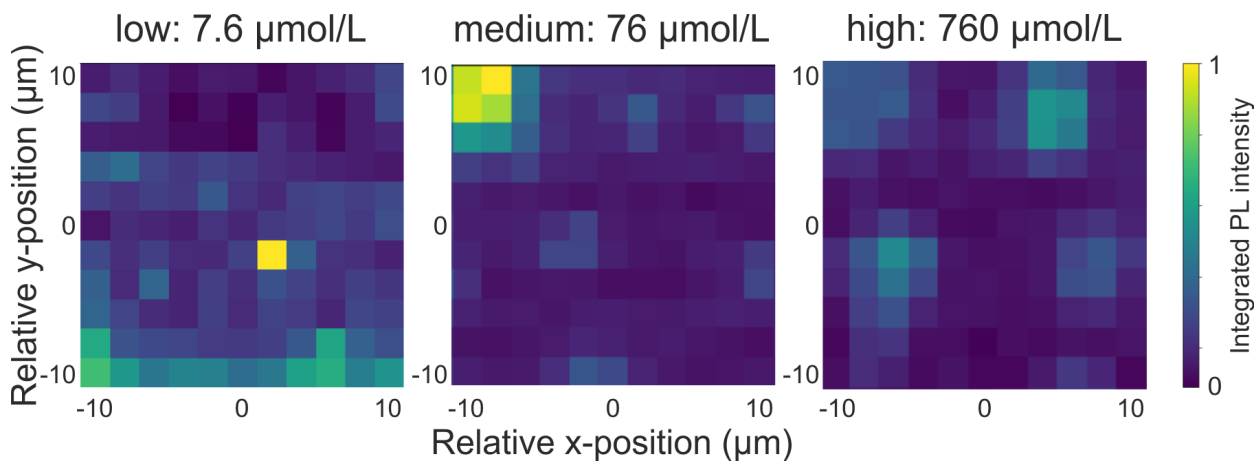

**Figure S11.** Comparison of the multilayer PL of three different concentration areas. For each concentration a typical PL map of the aged sample with a region of 400 $\mu$m² and a lateral step-size of 2 µm is displayed. The colloidal $WS_2$ nanosheets seem to agglomerate to larger features with increasing concentration.

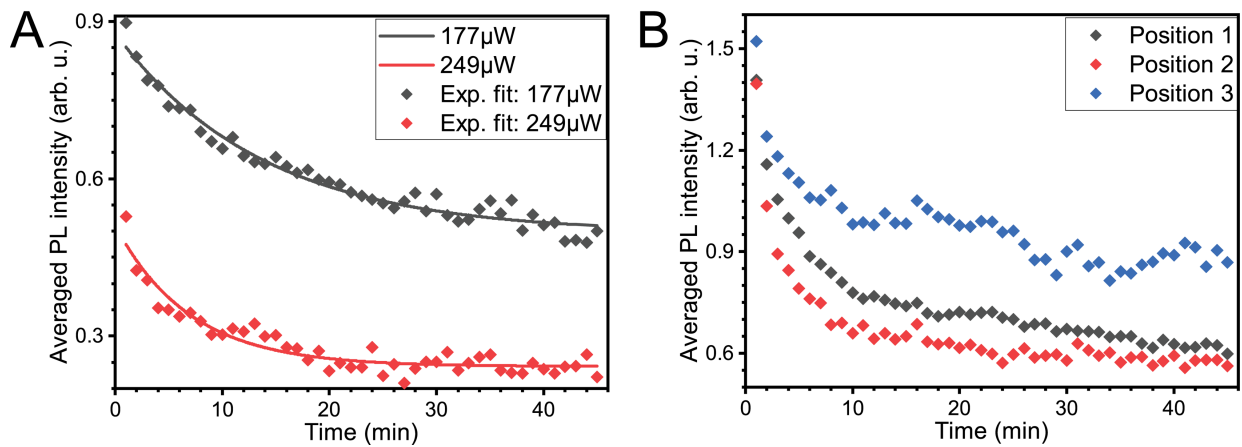

**Figure S12.** [A] Intensity dependence of the photodegradation process including an exponential fit of the integrated PL intensity. [B] Photodegradation process at different positions in the high concentration area using excitation powers of 170 $\mu$W.

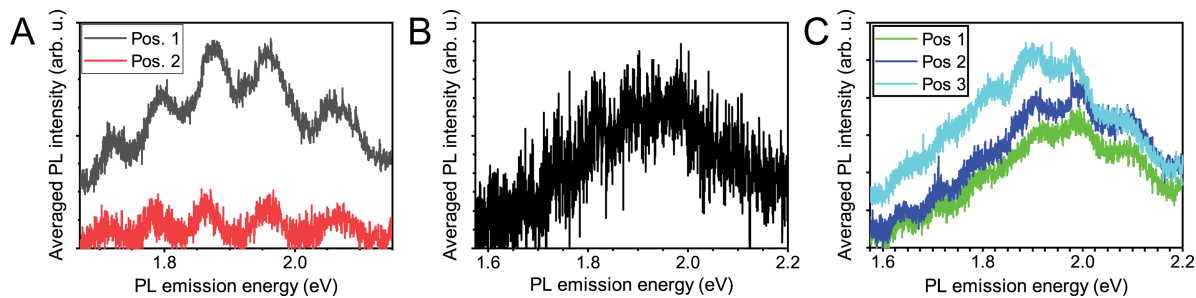

**Figure S13.** Characteristics of the interference effect: [A] PL spectra at different positions of the high concentration area of the aged sample, acquired with laser excitation powers of around 230 $\mu$W. [B] PL spectrum with a negligible interference effect acquired in the medium dilution area of the fresh sample 1 with a laser excitation power of 175 $\mu$W. [C] PL spectra at different positions of the high concentration area of the fresh sample 1, acquired with laser excitation powers of around 180 $\mu$W.

**Supplemental Tables:**

**Table S1. Parameter of the exponential fit of the photodegradation induced PL decrease.**

| Sample | Concentration | Excitation intensity ($\mu$W) | Lifetime $\tau$ (min) | PL intensity offset $I_0$ (arb. u.) |
|---|---|---|---|---|
| fresh | medium[a] | 178 | 13.7$\pm$1.2 | 0.496$\pm$0.01 |
| fresh | medium[a] | 249 | 6.9$\pm$0.8 | 0.242$\pm$0.005 |
| fresh | high[b] | 194 | 8.9$\pm$0.6 | 0.722$\pm$0.002 |
| fresh | high[c] | 174 | 5.5$\pm$0.4 | 0.662$\pm$0.008 |
| fresh | high[c] | 171 | 3$\pm$0.3 | 0.605$\pm$0.007 |
| fresh | high[c] | 173 | 7.5$\pm$1.3 | 0.894$\pm$0.014 |
| aged | high | 215 | 4.5$\pm$0.4 | 0.624$\pm$0.009 |
| aged | high | 200 | 4.6$\pm$0.4 | 0.306$\pm$0.007 |
| aged | high | 203 | 11$\pm$1 | 0.419$\pm$0.012 |

Related figures: [a]Figure S12A; [b]Figure 3E; [c]Figure S12B

**Table S2. Concentration-dependent multilayer PL characteristics of colloidal WS$_2$.**

| Sample | Calculated W-concentration (μmol/L) | Evaluated area (μm$^2$) | Number of multilayers per 100 μm$^2$ | Average PL intensity (min. /max.) |
|---|---|---|---|---|
| Sample 1 (fresh) | 7.6 | ~1200 | 1.1 | 21.62 (5.5/29.5) |
|  | 76 | ~10000 | 1.98 | 59.4 (4/451.5) |
|  | 760 | ~18300 | 1.07 | 51.4 (3.5/860) |
| Sample 2 (aged) | 7.6 | ~1200 | 3.03 | 29.3 (8/ 164.5) |
|  | 76 | ~5800 | 2.35 | 57.9 (8/ 276.5) |
|  | 760 | ~3800 | 1.9 | 95 (12.5/836.5) |

**Table S3. Parameters of the adapted Poisson distribution for the concentration dependence of the colloidal WS$_2$ multilayer PL.**

| Calculated tungsten concentration (μmol/L) | $r$ | $b$ | $I_0$ |
|---|---|---|---|
| 7.6 | 5.4 ± 0.9 | 0.25 ± 0.05 | 79.8 ± 12 |
| 76 | 5.5 ± 0.7 | 0.12 ± 0.02 | 26.1 ± 3 |
| 760 | 3.8 ± 0.5 | 0.053 ± 0.007 | 9.2 ± 1 |